# Data Collection through Vehicular Sensor Networks

*Abstract*— **N**ow a days Many car manufacturers are planning to install wireless connectivity equipment in their vehicles to enable communications with "roadside base station" and also between vehicles, for the purposes of safety, driving assistance, and entertainment. One distinct feature is that vehicles are highly mobile, with speed up to 30 m/s, though their mobility patterns are more predictable than those of nodes in Mobile Ad-hoc Networks (MANET) due to the constraints imposed by road, speed limits, and commuting habits. Therefore, these networks require specific solutions and identify a novel research area, i.e., Vehicular Ad-hoc Networks (VANET). In this paper, we focus on a particular VSN architecture, where the ad hoc network is operated by a telecommunication/service provider to combine non-valuable individual sensed data and extract from them effective feedbacks about the situation of the road in a geographical area. In operated VSNs, providers tend to reduce the traffic load on their network, using the free-frequency communication medium (IEEE 802.11p, for example). To do so, we propose TCDGP (Tree Clustered Data Gathering Protocol), a cross layer protocol based on hierarchical and geographical data collection, aggregation and dissemination mechanisms. We analyze the performances of our solution using a simulation environment and realistic mobility models. We demonstrate the feasibility of such solution and show that TCDGP offers the operator precious information without overloading his network.

## 1. INTRODUCTION

Over the past decade the nature of wireless communications has evolved rapidly. The introduction of 3G and WLAN technologies and the recent standardization of WiMax have helped to realize the vision of ubiquitous connectivity. Currently, much research effort is focusing on exploiting this "always-on" feature for use in Transportation Systems. The primary objective of ITS (Intelligent Transportation Systems) is to improve traffic safety, efficiency, and travelling comfort.

VSNs can be built on top of VANET by equipping vehicles with onboard sensing devices as shown in the above figure. Here, sensors can gather not only safety-related information, e.g., seat occupation, but also more complex multimedia data, e.g., video data. Unlike traditional sensor networks, VSNs are not subject to major memory, processing, storage, and energy limitations. However, the typical scale of a VSN over wide geographic areas (e.g., millions of nodes), the volume of generated data (e.g., streaming data), and mobility of vehicles make it infeasible to adopt traditional sensor network solutions where sensed data tends to be systematically delivered to sinks using data-centric protocols such as Directed Diffusion. Further, the mobility of sensor nodes makes it less efficient to use mobile agents, or MULEs in static sensor networks, which pick up data from sensors when in close range, buffer it, and drop off the data to wired access points. This project mainly focuses on developing efficient data gathering, searching protocols. Moreover, the goal includes test-bed implementation and middleware software development.

Besides DSRC, we can utilize cellular communications (2/3G) via Smartphone's. Recent Smartphone's are equipped with various sensors such as GPS, camera, audio, and video, and support various communications means such as 2/3G, Wi-Fi, and Bluetooth. Bluetooth enables us to connect other external sensors via a wireless data acquisition board. The particular importance of 2/3G connection is that it gives an always-on Internet connection, which makes data access and retrieval amenable.

In this work, we focus on the main component in Intelligent Transportation Systems, which is vehicular communication. Indeed, many car manufacturers are installing wireless connectivity equipments in their vehicles to enable communication between vehicles and also with the infrastructure. Vehicular Sensor Networks (VSNs) can be built on top of these vehicular networks by equipping vehicles with onboard sensing devices. In such case, sensors can gather a set of information like video data, speed, localization, acceleration, temperature, seat occupation, etc. Compared to traditional sensor networks, this recently emerged sensor network is not restricted by the power supply and the storage space. However, the typical scale of a VSN over wide geographic areas (e.g., millions of nodes), the volume of generated data (e.g., streaming data), and mobility of vehicles make it infeasible to adopt traditional sensor network solutions where sensed data tends to be systematically delivered to sinks using data-centric protocols such as Directed Diffusion [1].

Under such environment, an effective and efficient architecture for data collection and data exchange is more important. This work deals with such a system framework consisting of mobile vehicular sensors and road-side-units operated by an operator or a service provider (WiMax access point, 2.5/3G base station). Road-side-units are distributed over the road for collecting data from mobile vehicular sensors passing by. While mobile sensors on vehicles sense and send the information to road side units.

In our study, the VSN will be used by an infrastructure wireless network owner (telecom or Internet provider, for example) to gather "useless" individual information (that cannot be used to deduce a global view of the road traffic) from each vehicle and to aggregate them inside an ad hoc wireless network using free frequency (IEEE 802.11, for example), to get a global view of the state of the road in a geographical area at a specific time, or to use these information as a database for a posterior treatment (see figure1).

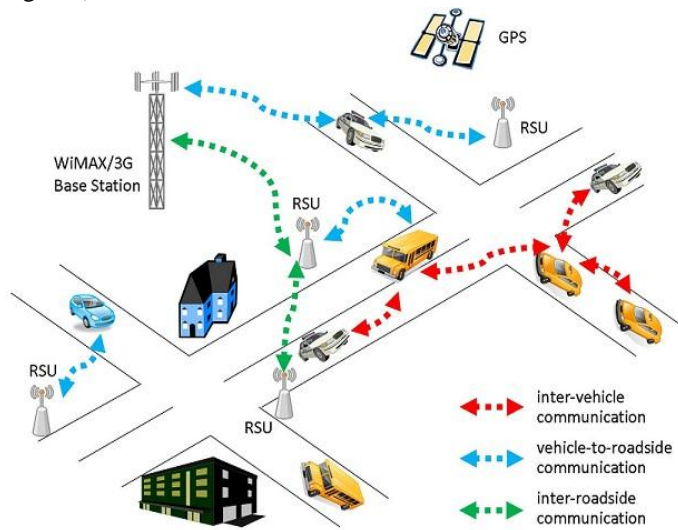

Figure 1 – Operated vehicular sensors network

In our study, the VSN will be used to collect individual information from each vehicle and aggregate them inside the ad hoc wireless network (free frequency). The aggregated information will be sent to the road side unit owned by the operator via a non-free frequency (WiMax or 2.5/3G). In fact, these sensors can generate massive amounts of sensed data and there is a need to collect, store, and retrieve them. The objective in such architectures for an operator/service provider is to reduce the use of its high-cost links. To do so, we present TCDGP (Tree Clustered Data Gathering Protocol): a cross-layered protocol based on hierarchical and geographical data gathering, aggregation and dissemination. The goal of CGP is to gather data from all nodes in the vehicular ad hoc network in order to offer different kind of ITS services. For example, a telecommunication/service provider can use CGP to provide:
- A real-time traffic information service, by gathering all node's positions and velocities, [2]
- A geographical localization service for customers who want to follow their vehicles mobility (fleet management),
- A parking lots availability service, by detecting empty spaces in parking lots, [3]
- Warnings messages in a specific area, when an unusual event happens (a sudden speed decrease of several vehicles, for example), [4]
- A real-time fuel consumption and pollution indicators, [5]
- Surveillance applications such as proposed in [2] where nodes make videos of the road and detect and save the registration plates of vehicles around.

## 2. INTRODUCTION ABOUT VEHICULAR SENSOR NETWORKS

VSNs come out as a new brand of vehicular networks, whose purpose is the real-time gathering and diffusion of information. In [5], the author used a VSN for a better understanding of the traffic jam formation. They pointed out the fact that vehicular sensor networks are one of the costless solutions which tends to reduce traffic jams, $CO_2$ emissions and fuel consumption. In [7], authors proposed the use of VSNs for security issues where agent nodes can look for a stolen car for example, by sending a query to all nodes that have crossed that vehicle. Another application of VSNs is the one proposed in [2] where the network provide the road users a more safety driving by disseminating alert messages in case of an emergency. A VSN can be considered as a fusion of a Vehicle ad hoc network (VANET) and wireless sensor network (WSN).

However, a VSN has some properties like:
- (i) Higher capacity since the inboard sensors is supplied with more energy, storage and computing capabilities comparing to well known sensor networks, for example.
- **(ii)** Huge amounts of data since a vehicle could be equipped by a lot of sensors (cameras, etc.).
- **(iii)** Dynamic data-sinks management since data sinks could be mobile compared to traditional WSNs, and
- **(iv)** Large scale connectivity since wide roads and grand avenues in urban environments can contain thousands of vehicles.

These specific characteristics have important implications for designing decisions in these networks. Thus, numerous research challenges (e.g. data dissemination, data aggregation, self-organization mechanisms) need to be addressed for vehicular sensor networks to be widely deployed.

### A. VSN Architectures

The data dissemination in vehicular sensors networks can be based upon three architectures as shown in figure 2:
- **V2V**: where both the collection and the restitution of information are done within the vehicular network. This solution may be used for quick alert messages dissemination for example.
- **V2I**: infrastructure based wireless links (GSM, UMTS, WiMax, Wifi/Mesh, etc.) are used to gather the data from VSN nodes.
- **Hybrid**: uses both V2V and V2I architectures.

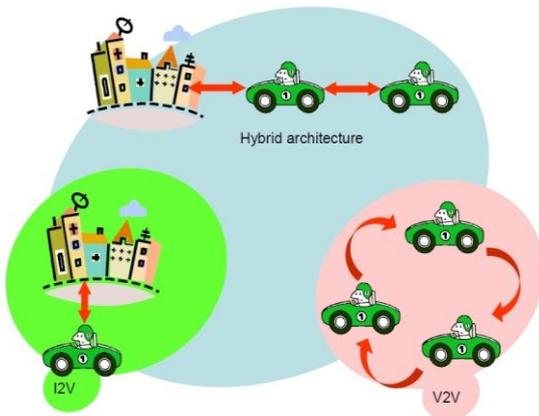

Figure 2- Vehicular communication architectures.

### B. Data Dissemination in VSN
We found in the literature different approaches for data dissemination in a VSN:

*1) Opportunistic dissemination*
Due to the intrinsic network partitioning of VSNs, some works, such as [6], recommend the use of opportunistic diffusion of data, in which messages are stored in each intermediate node and forwarded to every encountered node till the destination is reached. Thus the delivery ratio is improved. However, this kind of mechanisms is not suitable for non-tolerant delay applications. Opportunistic dissemination protocols have potentially applications in the domain of vehicular networking, ranging from advertising to emergency/traffic/parking information spreading: one of the characteristics of vehicular networks is that they are often partitioned due to lack of continuity in connectivity among cars or limited coverage of infestations in remote areas. Most available opportunistic, or delay tolerant, networking protocols, however, fail to take into account the peculiarities of vehicular networks.

*2) Geographical dissemination*
Due to the fact that end to end paths are not constantly present in a VSN, a geographic dissemination is used in [2] by sending the message to the closest node toward the destination till it reaches it. Another way to do geographic dissemination is given in [6] where the authors show how to use geo-casting to deliver messages to several nodes in a geographical area.

*3) Peer-to-peer dissemination*
In a P2P solution, the source node stores the data in its storage device and do not send them in the network till another node asks for them. In [2], such architecture is proposed for delay tolerant applications. peer-to-peer (P2P) systems such as BitTorrent [7], Slurpie [29], SplitStream [5], Bullet' [18] and Avalanche [10], to name but a few. The key idea is that the file is divided into M parts of equal size and that a given user may download any one of these (or, for network coding based systems such as Avalanche, linear combinations of the bits in the file parts) either from the server or from a peer who has previously downloaded it. That is, the end users collaborate by forming a P2P network of peers, so they can download from one another as well as from the server. Our motivation for revisiting the broadcasting problem is the performance analysis of such systems.

*4) Cluster-based dissemination*
For a better delivery ratio and to reduce broadcast storms, a message has to be relayed by a minimum of intermediate nodes to the destination. To do so, nodes are organized on a set of clusters, in which one node or more (Cluster Head) gathers data in his cluster and send them after to the next cluster. Cluster-based solutions provide less propagation delay and high delivery ratio with also bandwidth equity. In [4] the authors use a distributed clustering algorithm to create a virtual backbone that allows only some nodes to broadcast messages and thus, to reduce significantly broadcast storms.

We are interested in this paper, in the cluster based dissemination mechanisms combined with the geographical and opportunistic approaches.

### C. Data Aggregation
Data aggregation is a well known concept in Wireless Sensors Networks; it allows the nodes to merge, update or delete some information because they might be duplicated, similar or expired. There are several aggregation mechanisms proposed in the literature for sensors networks that can also be used in VSNs: In

[8], a timestamp aggregation technique is developed upon an opportunistic dissemination solution. In this case, if a node receives an information, it can decide if it is valid or not, by checking its sending time. Authors of [9] use a ratio-based and a cost-based algorithm to choose which information is important to aggregate and to estimate the error that can introduce a message in the data.

Another approach in data aggregation is introduced in [10]. The authors use Flajolet-Martin Sketch [11] to estimate similar data in a set of N entities and to merge these information. In this paper, we use a simple aggregation mechanism based on both timestamps and mean value calculation, where each set of nodes in a specific area will calculate an average value of collected data dropping some non-needed information (messages from another area, from other direction, etc.)

## 3. PROPOSED EFFICIENT PROTOCOL

As in the above-mentioned section, the energy efficiency in tree-based protocol like TREEPSI is better than cluster based and chain-based protocol. If some sensor nodes send data to the sink, this information of nodes will make a detour. Thus, that will cause more power dissipation in data gathering. This situation is happened as building the binary tree paths, especially when the sensor field is large and the numbers of sensor nodes are large. In order to improve the reduction of power dissipation, we propose a novel protocol to combine the cluster-based and tree-based protocol to improve it. In the following, we will describe the deployment and method of the protocol. And the first we can see the flow chart of protocol clearly as Figure 3. According to reference above-mentioned routing protocols, the network assumptions can be initiated as follows [4, 5, and 6].

1. Each node or sink has ability to transmit message to any other node and sink directly.
2. Each sensor node has radio power control node can tune the magnitude according to the transmission distance.
3. Each sensor node has the same initial power in WSNs.
4. Each sensor node has location information.
5. Every sensor nodes are fixed after they were deployed.
6. WSNs would not be maintained by humans.
7. Every sensor nodes have the same process and communication ability in WSNs, and they play the same role.
8. Wireless sensor nodes are deployed densely and randomly in sensor field.

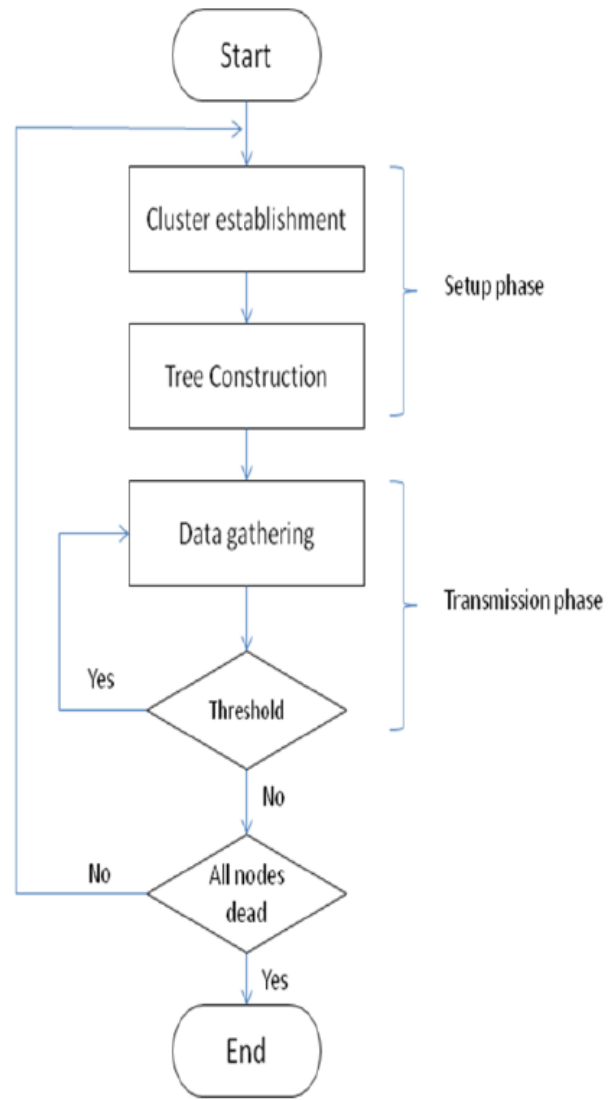

Fig 3 Flow chart for TCDGP

Sink could get the whole location and energy information about sensor nodes by two or other manners. One is recorded in the sink at the initial state as nodes were deployed. The other is that sink broadcast whole network, and then received the back message form sensor nodes.

### A. Cluster Establishment
Setup Phase:

This phase consists of two major steps: cluster formation and cluster head selection. Once the base station forms the primal clusters, they will not change much because all sensor nodes are immobile, whereas the selected cluster head in the same cluster may be different in each round. During the first round, the base station first splits the network into two sub clusters, and proceeds further by splitting the sub clusters into smaller clusters.

| Neighbor id | Residual Energy | Distance | Distance to BS | State | Weight |
|---|---|---|---|---|---|

The base station repeats the cluster splitting process until the desired number of clusters is attained. When the splitting algorithm is completed, the base station will select a cluster head for each cluster according to the location information of the nodes.

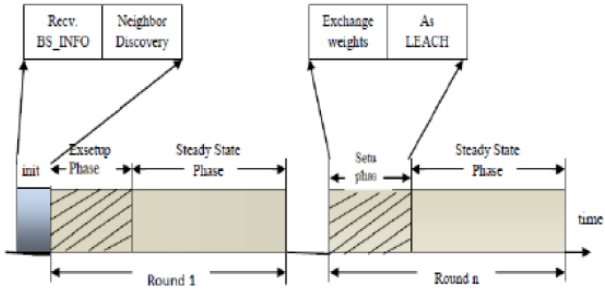

Figure-4: Extended Round Of Table

For a node to be a cluster head, it has to locate at the center of a cluster. Once a node is selected to be a cluster head, it broadcasts a message in the network and invites the other nodes to join its cluster. The other nodes will choose their own cluster heads and send join messages according to the power of the many received broadcast messages. When the cluster head receives the join message from its neighbor node, it assigns the node a time slot to transmit data. When the first round is over and the primal cluster topology is formed, the base station is no longer responsible for selecting the cluster head. The task of cluster formation is shifted from the base station to the sensor nodes. The decision to become a new cluster head is made locally within each cluster based on the node's weight value. The pseudo code for all operation is given below:

Initialize {
    1. Base station: acquire the number of clusters N;
    2. Split the network into N clusters;
    3. Choose cluster head from each cluster;
    4. Notify the node to be cluster head.
    }
Repeat:
{
1. Node i: if (Receive the notify message from the base station)
2. Work in cluster head mode;
3. If (Receive the broadcast message from cluster head node)
4. Work in sensing mode.
    }
For cluster head i:
{
    1. Receive data born cluster member j;
    2. Compute the weight value $W_i$ and $W_j$;
    3. If ($W_i > W_j$), $W_i$ Work in cluster head;
    4. Else i work in sensing mode;
    5. Notify j to be cluster head ;
    }

*B. Constructing Cluster Based Tree*

Sink will collect the information that cluster head had labeled in each cluster and build path in minimum spanning tree to compute the tree path. The Minimum Spanning tree (MST) concept in the Greedy algorithms used to solve the undirected weight graph problem. After eliminating some of the connection links, the sub-graph still have the connection ability. For this reason, sub-graph can reduce the sum of the weights. A sub-graph who has the minimum sum of weights must be a tree like framework. Spanning tree could let all nodes conform to tree definition which is connected in the graph. A connected sub-graph which has a minimum sum of weights must be a spanning tree. On the contrary, it is not correctly absolutely. There could be several kinds Minimum spanning tree in a graph, and it is not the only one. But their sum of weight should be the same. If we use Brute Force to find the minimum spanning tree, it will produce huge computation time. In order to avoid this, we use Prim algorithm to help us finding the MST.

*C. Data Aggregation*

After the routing mechanism has established, every tip nodes transmit gathering data to upper level nodes. Then the upper level nodes will fuse received data and sensed data by itself, and send the data to next upper level nodes. The process will keep going until the root node, cluster head, has aggregated the data in the cluster. It is called a round‖ as all root nodes has finished transmitting data.

## 4. PERFORMANCES EVALUATION

To validate and evaluate TCDGP, we have chosen Qualnet 4.5 simulation environment. We also extended and adapted the mobility model proposed in [11] to our needs. Our tool generates realistic random vehicles' displacements.

*A. Assumptions*

*1) Spatiotemporal environment*

We execute TCDGP on a straight road section partitioned into 18 equal segments, as depicted in Figure 9. The base station that covers all the section is present at one end point of the road. All the key parameters of our simulation are summarized in the following table:

| SIMULATION / SCENARIO | | MAC / CGP | |
|---|---|---|---|
| Simulation time | 600s | MAC protocol | 802.11b |
| Map size | 2500x2500 m2 | Capacity | 2 Mbps |
| Mobility model | VanetMobisim [12] | Trans. Range | ~266 m |
| Number of seg. | 18 | HL_PT | ~0.1 s |
| Nodes | 50 - 1000 | PK_PT | ~0.2 s |
| Vehicle velocity | 0 – 108 km/h | IS_PT | ~0.1 s |
| Segment length | 100 m | FULL_DURATION | 5 s |
| Road length | 1.8 km | CHD_DURATION | 1 s |
| Road width | 15 m | GAT_DURATION | 3 s |
| Number of lanes | 2 | AGG_DURATION | ~0.1 s |
| Store, carry and forward | Not used | DIS_DURATION | 1 s |

Table 1: Simulation Setup

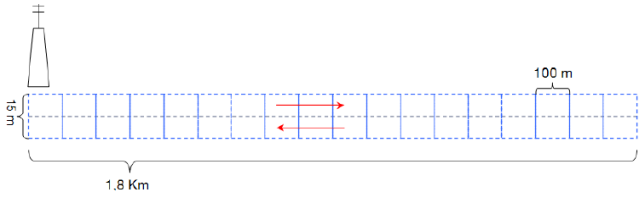

Fig.5 Spatial Environment

### B. Simulation Scenarios

*1) Scenario 1: Per node dissemination*

In this scenario, each node sends its collected data (speed, position, etc.) individually and periodically to the base station using the provider's cellular network. The aggregation in this case, is done at the Telco provider level. (See Figure 6)

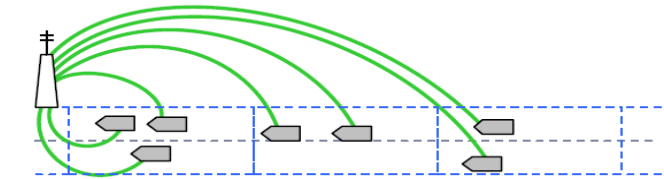

Figure 6 – Per node dissemination scenario.

*2) Scenario 2: Per Cluster Head dissemination*

In this scenario (see Figure 7), the local data gathering and aggregation are done at the segment level, as described in TCDGP. The aggregated data (average speed, number of nodes, etc.) are sent to the base station directly from the cluster head of each segment. The Telco provider will only aggregate the data from each segment.

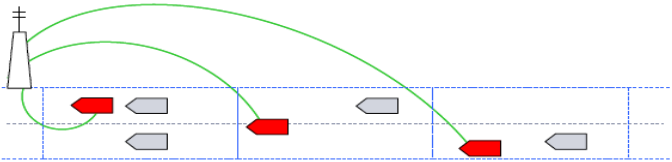

Figure 7 – Per cluster head dissemination scenario.

*3) Scenario 3: Complete TCDGP dissemination*

As depicted in Figure 8, TCDGP will be integrally executed in this scenario, from the cluster head election to the data dissemination to the provider.

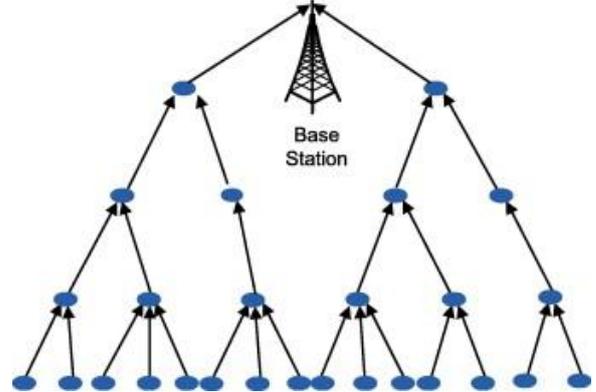

Figure 8- Complete TCDGP dissemination

### C. Simulation Results

We calculate the number of messages sent to the base station via the provider's cellular network during 600 seconds.

Thus, we can see in which scenario the data collection is the greediest in terms of cellular network usability.

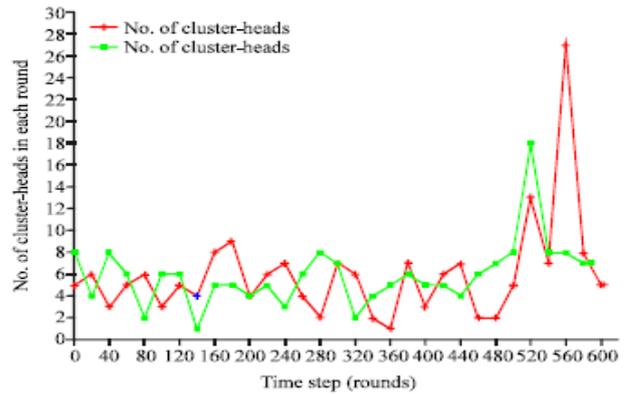

Figure 9 – Numbers of V2I messages

## 5. CONCLUSION

In this paper, a novel data gathering and dissemination system (TCDGP, Tree Clustered Data Gathering Protocol) based on hierarchical and geographical dissemination mechanisms on vehicular sensors networks is proposed. Designed for hybrid VANET architecture, it allows telecommunication/service providers to get valuable information about the road environment in a specific geographical area, using V2V network to minimize the high-cost links usability and base stations to gather information from the vehicles. Simulations results of TCDGP demonstrate the feasibility of the proposed approach; moreover, they show that TCDGP reduces considerably the provider's network usability without any loss of accuracy in the collected data. We are currently extending this work by performing other extensive simulation in order to study all the TCDGP parameters.